\documentclass[times,twocolumn,final,authoryear]{elsarticle}

\usepackage{jasr}
\usepackage{framed,multirow}

\usepackage{amssymb}
\usepackage{latexsym}

\usepackage{url}
\usepackage{xcolor}
\definecolor{newcolor}{rgb}{.8,.349,.1}

\usepackage{aas_macros}
\newcommand{\nuo}{{\nu_{\rm{o}}}}

\usepackage[citebordercolor=white]{hyperref}

\journal{Advances in Space Research}

\begin{document}

\verso{Vladim\'{\i}r Karas et al}

\begin{frontmatter}

\title{Electromagnetic signatures of strong-field gravity from accreting black~holes\tnoteref{tnote1}}\tnotetext[tnote1]{Based on a contribution solicited at 43$^{\rm rd}$ COSPAR Scientific Assembly, Sect. E1.4 ``Black Hole Astrophysics: Observational Evidence and Theoretical Models'' (28 January -- 4 February 2021).}

\author[1]{Vladim\'{\i}r \snm{Karas}\texorpdfstring{\corref{cor1}}{}}
\cortext[cor1]{Corresponding author: V.K. Email address: vladimir.karas@asu.cas.cz}
\author[2,3]{Michal \snm{Zaja\v{c}ek}}
\author[4]{Devaky \snm{Kunneriath}}
\author[1]{Michal \snm{Dov\v{c}iak}}

\address[1]{Astronomical Institute, Czech Academy of Sciences, Bo\v{c}n\'{\i} II 1401, CZ-141\,00 Prague, Czech Republic}
\address[2]{Center for Theoretical Physics, Polish Academy of Sciences, Al.\ Lotnikow 32/46, P-02\,668 Warsaw, Poland}
\address[3]{Department of Theoretical Physics and Astrophysics, Faculty of Science, Masaryk University, Kotl\'a\v{r}sk\'a 2, Brno, CZ-611\,37, Czech Republic}
\address[4]{National Radio Astronomy Observatory, Charlottesville, VA 22903, USA}

\received{20 March 2021}
\finalform{5 September 2021}
\accepted{12 September 2021}
\availableonline{22 September 2021}

\begin{abstract}
Observations of galactic nuclei help us to test General Relativity. Whereas the No-hair Theorem states that classical, isolated black holes eventually settle to a stationary state that can be characterized by a small number of parameters, cosmic black holes are neither isolated nor steady. Instead, they interact with the environment and evolve on vastly different time-scales. Therefore, the astrophysically realistic models require more parameters, and their values likely change in time. New techniques are needed in order to allow us to obtain independent constraints on these additional parameters. In this context, non-electromagnetic messengers have emerged and a variety of novel electromagnetic observations is going to supplement traditional techniques in the near future. In this outline, we summarize several fruitful aspects of electromagnetic signatures from accretion disks in strong-gravity regime in the outlook of upcoming satellite missions and ground-based telescopes. As an interesting example, we mention a purely geometrical effect of polarization angle changes upon light propagation, which occurs near the black hole event horizon. Despite that only numerical simulations can capture the accretion process in a realistic manner, simplified toy-models and semi-analytical estimates are useful to understand complicated effects of strong gravity near the event horizon of a rotating black hole, and especially within the plunging region below the innermost stable circular orbit.
\end{abstract}
\end{frontmatter}


\section{Introduction}
\label{sec1}
Multitude of evidence supports the most intriguing revelation in astronomy: dark, compact objects in cores of galaxies \citep{1994ASIC..445.....G,1999agnc.book.....K,2006eac..book.....S}. They have been looked for via a broad variety of independent observational approaches that have gathered multi-wavelength evidence about strong gravity effects in radiation across the entire electromagnetic spectrum -- namely, the case for a rapid motion of gas and stars, large Doppler and gravitational shifts of spectral lines, occurrence of jets, rapid variability of light-curves and other signatures -- thought to provide persuasive hints for the existence of supermassive black holes (SMBHs) in active galactic nuclei (AGN) and quasars \citep{1986qa...book.....W,2009elu..book..138P}, as well as centres of non-active galaxies including the Milky Way \citep[Sagittarius A*;][]{2005bhcm.book.....E,2007gsbh.book.....M,2010RvMP...82.3121G,2017FoPh...47..553E,2021arXiv210213000G}. The specific features are testable and the modeling of spectra and light-curves captures many properties of present-day observations \citep[e.g.][and further references cited therein]{2003PhR...377..389R,2014A&ARv..22...72U}. Accretion is often accompanied by ejection of a significant fraction of material, and the energy of emerging radiation spans from radio over X-rays to gamma radiation \citep{1994IAUS..159.....C,2005Ap&SS.297..131C,2017FrASS...4...35P}. Recently, radiation signatures have been supplemented by multiple (non-electromagnetic) messengers: gravitational waves which are detected from mergers of stellar black holes \citep{PhysRevLett.116.061102}, and ultra-high energy particles of cosmic rays including high-energy neutrinos emerging presumably from inner jets near SMBHs in blazars \citep{2017muas.book.....B,2019NatRP...1..585M,2019NewAR..8601525D,2019A&A...630A.103B,2021arXiv210203366R,2021arXiv210300292B} as well as in radio-emitting tidal disruption events \citep{2021NatAs...5..510S}.

According to General Relativity (GR), black holes are defined by their event horizon and they are described by the spacetime metric \citep[GR;][]{1984ucp..book.....W,1998mtbh.book.....C}. The standard scenario requires just two parameters to describe uniquely the spacetime of an isolated, stationary black hole: mass $M_\bullet$ and angular momentum $J_\bullet$ \citep{1967PhRv..164.1776I,1971PhRvL..26..331C,2016CQGra..33f4001B}. According to the cosmic censorship hypothesis, the singularity is thought to be hidden by the event horizon; it does not influence distant observers \citep{1969NCimR...1..252P,2002GReGr..34.1141P}. More parameters, such as the electric and magnetic charges, or the cosmological constant are mathematically possible but they are usually considered unimportant from the astrophysical point of view \citep{2010ApJ...722.1240K,2010CQGra..27m5006K,2021CQGra..38m5002M}. However, see \citet{2018MNRAS.480.4408Z}, and also \citet{2014PhRvL.112y1102S,2019PhRvR...1c3106B,PhysRevD.103.L021502}, where the possibilities of additional ``hairs'' were studied (this reaches beyond the classical GR and we will not pursue the idea further here). The situation becomes more complex if the black hole is surrounded by a significant amount of matter, for example in the form of a self-gravitating torus made of dust and gas, or a dense cluster of stars which can contribute as a source of the gravitational field \citep{2004CQGra..21R...1K,2010RvMP...82.3121G,2014A&A...566A..47S}.  Also, if the system lacks symmetries, the gravitational waves are emitted and the spacetime evolves in time \citep{2005LRR.....8....8M,2017PhRvD..96l3003K}. 

In the present contribution, we concentrate on the electromagnetic signatures of strong gravity near the event horizon of SMBHs. By this we traditionally mean the specific properties of light-curves and spectra from sources moving very rapidly at velocity reaching a significant fraction of the speed of light $c$ near the event horizon \citep{1972ApJ...173L.137C,1973ApJ...183..237C}. In these two seminal papers, the authors predicted periodic variations of the observed radiation flux that arrives from a star revolving around a fast-rotating SMBH to a distant observer. Their method of calculation provides a basis for very practical computational algorithms that have been developed to constrain parameters of the orbit and of the black hole. A number of  authors formulated the origin of light signal in terms of bright spots (point-like, extended circular, or elongated and time evolving), magnetic flares and loops (equatorial or levitating above the disk plane), and extended spiral waves and disturbances propagating across the accretion disk and radiating predominantly in the X-ray spectral domain \citep{2004fxra.book.....F}. The scenario was later generalized to the case of eccentric orbits that can be inclined at an arbitrary angle with respect to the black hole equatorial plane, and to arbitrary values of the spin \citep{1994ApJ...422..208K,2010MNRAS.402.1614D,2016MNRAS.457.1145P}. 

Nowadays, a technologically novel addition is the possibility of exploring the polarimetrical properties of light in X-rays, which often dominate the radiation signal from the black hole vicinity \citep{2010xpnw.book.....B,2014axp..book.....F}. Despite the fact that theoretical understanding of the light propagation is practically complete in terms of geometrical optics \citep[bending of light rays and changes of photon energy due to gravity;][]{1975ApJ...202..788C,1992MNRAS.259..569K,2004ApJS..153..205D,2006ApJ...651.1031S} and the associated wave propagation \citep[mathematically equivalent description of the deformation of wave fronts;][]{1977PhRvD..16..933H,Perlick2004}, more effort still needs to be devoted to actually {\em detect} the strong-gravity implications in the signal from various cosmic sources and use them to measure the parameters of the system in astrophysically realistic models \citep{2002apsp.conf..177B,2021cosp...43E1412R}. Let us note that (semi)analytical methods and various toy-models maintain their value along with modern realistic schemes, that are currently possible thanks to powerful computational resources. To date it appears unlikely that either of the approaches alone will enable a reliable comparison of the simulated results with the observational data on different sources. To that end the parameter space of the realistic models is far too large and the degeneracy between parameters is far too frequent. Nonetheless, fitting the observational datasets can constrain the physical properties of the plasma environment, geometry of different components of the sources, and also the parameters of the central black hole.

Despite the progress achieved during the recent decade, there are several directions where, even nowadays, the theoretical framework is not yet complete, neither it has been tested observationally. First, the propagation of light through non-vacuum (plasma filled) curved spacetime, where the primary photons interact with matter and the effects of strong gravity combine with those of a dispersive medium. Gravitational light bending \citep[independent of wavelength in GR but potentially wavelength-dependent in alternative theories of gravity;][]{2014LRR....17....4W,2016PhRvL.117i1101J,2016CQGra..33f4001B} then combines with refraction of light in the plasma (wavelength-dependent effect according to the dispersion properties of the medium). The theoretical description has not yet been fully developed in strong gravity, however, the essential aspects are well-known and an update was recently cast in the framework of current observational capabilities \citep{2017PhRvD..95j4003P,li2021gravitational}. Second, the role of strong gravity in changing the polarization properties of light, have only recently started to be confronted with observations \citep{2018NatAs...2...50V}. Light is strongly polarized if the synchrotron emission mechanism operates due to magnetic fields present in the accreting medium and emanating jets, but also reflection and reprocessing processes are capable of polarizing the emergent signal in non-spherical geometry of accretion disks. Polarization can reveal much about the intervening medium as about the source. It was noted by \citet{2009ApJ...701.1175S} that the polarization of photons, which return back onto the disk due to light bending, can be significantly enhanced \citep[see also][]{2020MNRAS.493.4960T,2020arXiv200615838R,2020MNRAS.498.3302W}. On the other hand, the impact of this effect is somewhat limited because the higher-order (strongly bent) photon rays produce the radiation flux exponentially weaker compared to the direct rays. Also, \citet{2009ApJ...703..569D} pointed out that the photons travelling at an oblique direction at infinity can be emitted at fluid-frame directions closer to the polar axis. The parallel transport of polarization then dilutes the observed polarization. This effect is strongest at small radii, and so the polarization is reduced toward higher energy.

As light from an accretion disk becomes polarized by scattering that occurs on electrons before the photon can escape to the outer space, the polarization degree is expected to be only moderate; to a very good precision, the emergent signal follows Chandrasekhar's theory for optically thick medium in the diffusion limit \citep[$\tau\gg1$;][]{1960ratr.book.....C}. The theoretical possibility of {\em testing\/} General Relativity via measuring spectral and polarimetric properties of the accreting black holes is exciting \citep{2016CQGra..33k3001J}; however, in the near future this approach will most likely produce only tentative results because the photon-hungry polarimetry can reach only very limited signal-to-noise ratio especially in X-rays \citep[e.g.][however, see further below for recent advances with Event Horizon Telescope in the millimeter regime]{2012ApJ...754..133K,2015EPJC...75..383L,2021ApJ...912...35N}. For comprehensive reviews on testing black hole candidate objects with electromagnetic radiation, see \citet{2017RvMP...89b5001B,2021SSRv..217...65B}. A more general overview of the outlook for evidence from other messengers including the gravitational waves can be found in \citet{2015CQGra..32x3001B,2019LRR....22....4C}, and further references cited therein.

Additional parameters are expected in the framework of generalized theories of gravity; there is an interesting promise to test the alternative theories by constraining the parameters that describe a departure from GR \citep[][and further references therein]{2010ApJ...718..446J,2015CQGra..32x3001B,2016CQGra..33e4001Y,2017RvMP...89b5001B,2020PhRvL.125n1104P}. However, as mentioned above, these alternative scenarios will require more precise observations than what is currently accessible \citep{2018GReGr..50..100K}. Systematic uncertainties and statistical errors of the measurement are still too large compared to the rather small observational differences between the viable spacetime models of astrophysical black holes.

The supermassive black hole in the center of our Galaxy is a unique target. At the distance of $\simeq8.3$ kpc, the core of Milky Way is one hundred times closer than the second nearest to us, which is a nucleus of a similar galaxy (M31 in the constellation of Andromeda). Given the large mass of about four million solar masses, $M_\bullet\simeq4 \times 10^6M_\odot$, Sgr~A* subtends the largest angular size on the sky of all known SMBHs, about $50\mu$arcsec (corresponding to the linear scale of about $0.1$ AU). As a convenient mass-scaling relation, we can introduce the fiducial mass, $M_7\equiv M_\bullet/(10^7M_\odot)$. The next largest black hole shadow is cast by SMBH in the core of M87 giant elliptical galaxy \citep[$M_7\simeq10^2$;][]{2019ApJ...875L...1E,2020ApJ...901...67W}. The dynamical time-scale and the event horizon radius are thus increased proportionally to the black hole mass. Despite the larger mass of M87 compared to Sgr~A*, this object is more distant from us and the accretion flow is accompanied by a bright jet that appears to be pre-accelerated and collimated at the very central region, with the diameter of the jet launching site \citep{2018A&A...616A.188K}. The emission area contains an unresolved superposition of the signal from both disk-like inflow and the jet ejection. 

We outline several effects of strong-gravity in the context of radiation signal from accreting black holes at different wavelengths. We provide selected useful references to the cornerstone original works as well as more recent papers, although it would be beyond the scope and extent of the brief summary to give the readers a comprehensive list.

\section{Light rays and wave fronts near a rotating black hole}
We describe the gravitational field of an astrophysical black hole by Kerr geometry of spacetime in General Relativity \citep{1963PhRvL..11..237K}. Self-gravitation of the accreted gas is not taken into account (the black hole is assumed to be isolated as far as the source of the gravitational field is concerned). The line element can be written in Boyer-Lindquist spheroidal coordinates in the form
\citep{2017grav.book.....M}
\begin{eqnarray}
ds^2&=&-\frac{\Delta}{\Sigma}\Big(dt-a\sin^2\theta\;d\phi\Big)^2
+\frac{\Sigma}{\Delta}\;dr^2+\Sigma\;d\theta^2
\nonumber \\
&&+\frac{\sin^2\theta}{\Sigma}\Big[a\;dt-\big(r^2+a^2\big)\;d\phi\Big]^2,
\label{eq:metric}
\end{eqnarray}
where the metric functions $\Delta(r)$ and $\Sigma(r,\theta)$ are known in an analytical form; the metric coefficients depend only on spheroidal radius $r$ and latitudinal angle $\theta$, and so the system is axially symmetric and stationary (manifestly independent of azimuth $\phi$ and time $t$). The event horizon exists for the dimensionless spin parameter $-1\leq a\equiv cJ_\bullet/GM_\bullet^2 \leq1$. The outer radius of the event horizon is located at $r_+=1+(1-a^2)^{1/2}$ and it hides the singularity from a distant observer; $\Delta(r_+)=0$ at the horizon. As soon as the black hole rotates (i.e., $a\neq0$), all particles and photons are forced to co-rotation (positive values of spin correspond to co-rotating, i.e.\ prograde motion, while negative values describe the retrograde sense of rotation). Lengths can be expressed in units of the gravitational radius, $R_{\rm g}{\simeq}GM_{\bullet}/c^2{\doteq}1.48\times10^{12}M_7$~cm. We can then employ geometrized units with $c=GM_{\bullet}=1$, which means that we scale lengths with the mass of the black hole; therefore, all quantities are dimensionless.

Particles can circle the black hole in the equatorial plane, but only if the radius of the orbit is above the marginally stable orbit $r=R_{\rm ms}(a)$ for the corresponding angular momentum \citep{1972ApJ...178..347B}. The orbital period of matter revolving along circular trajectory is $t_{\rm{}orb}=310(r^\frac{3}{2}+a)M_7$ seconds. It should be noted that the light-crossing time across $R_{\rm g}$ is also proportional to the black hole mass because these characteristic scales are defined solely by the Kerr geometry. 
It is interesting to note that spherical orbits are also possible outside the equatorial plane, however, Lense-Thirring precession occures \citep{1972PhRvD...5..814W,1994ApJ...422..208K}. Inclined accretion disks are thus necessarily twisted and the emerging signal is time dependent  \citep{1975ApJ...195L..65B,2010MNRAS.402..537W}.

The standard model of planar, geometrically thin accretion disks is based on the phenomenological $\alpha$-prescription of viscosity, which was originally introduced by \citet{1973A&A....24..337S}, in the relativistic version by \citet{1973blho.conf..343N,1974ApJ...191..499P}. Locally dissipated energy and the emerging radiation spectra can be computed \citep{2001ApJ...559..680H,2005ApJ...621..372D}. Matter in standard accretion disks resides in the equatorial plane; in the region of stable motion it follows strictly Keplerian circular orbits, the situation which we also assume in this paper. More general models of slim accretion disks rotate with an angular momentum visibly different from the Keplerian value \citep{2013LRR....16....1A}. The higher the accretion rate, the more significant the departure. Depending on the assumption about the accretion rate (provided at the outer boundary of the accretion disk), the mechanism of radial transport of matter, and about the inner boundary condition, accretion disks may or may not form an inner rim at or around the radius of the innermost stable circular orbit. In other words, in the strong-gravitational field regime close to the black hole, as long as the accretion rate is small (sub-Eddington), the stream lines of an accretion disk at each radius are well reproduced by the slender torus limit \citep{1985MNRAS.216..553B,2006MNRAS.369.1235B}, where the pressure terms are also small. On the other hand, enhancing the accretion rate, including the magnetic fields, and considering non-axisymmetric instabilities lead to the development of vertically extended coronae and outflows, where the emerging photons can be further reprocessed. The radiation reprocessing generally (with some notable exceptions) tends to diminish the basic signatures of strong gravity; however, these effects are beyond the scope of the present discussion. The latter has attracted renewed interest in the context of EHT measurements with the polarimetric resolution \citep{2021ApJ...910L..12E,2021ApJ...912...35N}.

The notion of the marginally stable orbit refers to a turnover in the effective potential of a particle motion, which implies that $R_{\rm ms}$ is the smallest radius at which stable circular trajectories are possible. It has become customary to designate this radius as the Innermost Stable Circular Orbit (ISCO; this term has been used mainly in astronomy oriented literature, while the historically original denomination of the marginally stable orbit puts more emphasis on the physical significance of that radius). The plunging region extends all the way from $R_{\rm ms}$ down to $r_+$. In that area particles cannot stay orbiting at constant radius; instead, they inspiral inward while maintaining the angular momentum of the ISCO. For a co-rotating equatorial motion the ISCO radius is given by \citep{1972ApJ...178..347B}
\begin{equation}
R_{\rm ms} = 3+z_2-\big[\left(3-z_1)(3+z_1+2z_2\right)\big]^\frac{1}{2},
\label{velocity1}
\end{equation}
where $z_1 = 1+\alpha_+\alpha_-[\alpha_++\alpha_-]$, $\alpha_{\pm}=(1{\pm}a)^\frac{1}{3}$, and $z_2 = (3a^2+Z_1^2)^\frac{1}{2}$. The ISCO spans the range from the dimensionless $R_{\rm ms}=1$ (case of maximally co-rotating black hole with the spin parameter $a=1$) to $R_{\rm ms}=6$ (a static black hole, $a=0$). 

A well-established scenario suggests that only very few spectral features originate in the plunging region, mainly because the density of the accreted medium drops while its ionization grows to high levels \citep[see][]{2014MNRAS.439.2307F}. There have been several arguments in favour of exploring this region in more detail especially if the medium is magnetized \citep{2005ApJ...622.1008K,2010A&A...521A..15A,2011ApJ...743..115N,2012MNRAS.424.2504Z}. The plunging region is directly adjacent to the event horizon, and so the GR effects on photons originating here should be most prominent, assuming that the signal from that region can be detected and disentangled \citep{1996AcHPh..69...69S,2018EPJC...78..180S,2021PhRvD.103d4029O}. We will show hereafter that polarimetrical effects contain even more specific signatures of strong gravity if they arise in the plunging area \citep{2006AN....327..961K,2009ApJ...701.1175S,2010ApJ...712..908S}.

\begin{figure}[tbh!]
\centering
\includegraphics[width=0.22\textwidth]{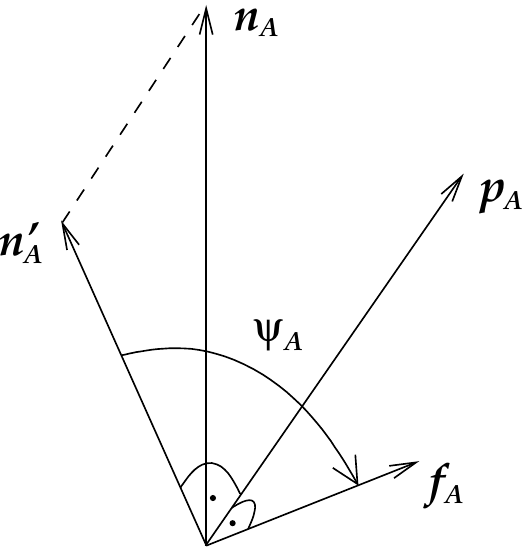}
\caption{On the definition of the change of polarization angle $\Psi$. (i) Let three-vectors $\vec{p_A}$, $\vec{n_A}$, $\vec{n'_A}$ and $\vec{f_A}$ be the momentum of a photon, normal to the disk, projection of the normal to the plane perpendicular to the momentum and a vector that is parallelly transported along the geodesic, respectively; (ii) let $\Psi_A$ be an angle between $\vec{n'_A}$ and $\vec{f_A}$; (iii) let the quantities in (i) and (ii) be evaluated at the disk for $A=1$ with respect to the local rest frame co-moving with the disk, and at infinity for \hbox{$A=2$} with respect to a stationary observer. Then the change of polarization angle can be defined as $\Psi=\Psi_2-\Psi_1$ \citep{2005ragt.meet...47D}.}
\label{fig0}
\end{figure}

The change of photon energy (gravitational and Doppler), bending of light rays, which implies the gravitational lensing (or de-focusing, in the dependence on geometrical settings), and mutual time delays along the photon path are the principal effects that arise due to or are significantly influenced by GR. Light propagation through a curved  space-time obeys the lensing relation, $F_{\!_{A,\,\rm em}}\,dS_{_{\rm em}}=F_{\!_{A,\,\rm o}}\,dS_{_{\rm o}}$ \citep{1992grle.book.....S}. In order to describe the energy changes, one can define the redshift $z$ at the point of emission and connect the point of emission with a distant observer,
\begin{equation}
1+z=\frac{(k_\alpha u^\alpha)_{\rm em}}{(k_\alpha u^\alpha)_{\rm o}},\quad
S_{\!_A}=\frac{k_{\!_A}}{(1+z)^2dS},
\label{eq:prop}
\end{equation}
where $k$ is the wave vector, $u$ is four-velocity of the source and the observer at points of light emission (em) and the point of observation (obs), respectively, and $dS$ stands for the cross-section defined by neighbouring light rays, which describes the lensing effect.

\citet{1975MNRAS.171..457R} proposed that polarized light can serve as an important source of additional information about accreting black holes in X-ray binaries. Four quantities $S_X$ can be directly connected to the traditional definition of the Stokes parameters \citep{1960ratr.book.....C,2014axp..book.....F}.
Normalized Stokes parameters can be introduced \citep{2005ragt.meet...47D},
\begin{equation}
i_\nu\equiv \frac{I_{\nu}}{E}\, ,\quad q_\nu\equiv \frac{Q_{\nu}}{E}\, ,\quad
u_\nu\equiv \frac{U_{\nu}}{E}\, ,\quad v_\nu\equiv \frac{V_{\nu}}{E}\, ,
\end{equation}
where $I_{\nu}$, $Q_{\nu}$, $U_{\nu}$ and $V_{\nu}$ are Stokes parameters for light with frequency $\nu$ ($E=h\nu$ is the corresponding energy). The detector observables are Stokes parameters per energy bin, $\Delta i_{\rm o}$, $\Delta q_{\rm o}$, $\Delta u_{\rm o}$ and $\Delta v_{\rm o}$, which are related to the local (intrinsic) values $i_{\rm loc}$, $q_{\rm loc}$, $u_{\rm loc}$, and $v_{\rm loc}$ on the disk plane:
\begin{eqnarray}
\label{S1}
{\Delta}i_{\rm o}(E,\Delta E) & = & X(i_{\rm loc}\,F)\, ,\\
\label{S2}
{\Delta}q_{\rm o}(E,\Delta E) & = & X\big([q_{\rm loc}\cos{2\Psi}-u_{\rm loc}\sin{2\Psi}]\,F\big)\, ,\\
\label{S3}
{\Delta}u_{\rm o}(E,\Delta E) & = & X\big([q_{\rm loc}\sin{2\Psi}+u_{\rm loc}\cos{2\Psi}]\,F\big)\, ,\\
\label{S4}
{\Delta}v_{\rm o}(E,\Delta E) & = & X\big(v_{\rm loc}\,F\big)\, .
\end{eqnarray}
Here, $X(f)\equiv N_0\int{\rm d}S\,\int{\rm d}E_{\rm loc}f$, $F\equiv F(r,\varphi)=g^2\,l\,\mu_{\rm e}$ are transfer functions, $\Psi$ is the angle by which polarization vector rotates while it is transported along the null geodesics.

The change of polarization angle $\tan{\Psi}={Y}/{X}$ is defined by \citep{1977Natur.269..128C,1980ApJ...235..224C}
\begin{eqnarray}
X & = & -(\alpha-a\sin{\theta_{\rm o}})\kappa_1-\beta\kappa_2\, ,\\
Y & = & \phantom{-}(\alpha-a\sin{\theta_{\rm o}})\kappa_2-\beta\kappa_1\, ,
\end{eqnarray}
with $\kappa_{1,2}$ being components of the complex constant of motion $\kappa_{\rm pw}$, $\alpha$ and $\beta$ are impact parameters \citep[see][]{1970CMaPh..18..265W}. See Figure~\ref{fig0} for the definition of the angles.

\begin{figure}[tbh!]
\vspace*{-7em}
  \centering
  \includegraphics[scale=0.45]{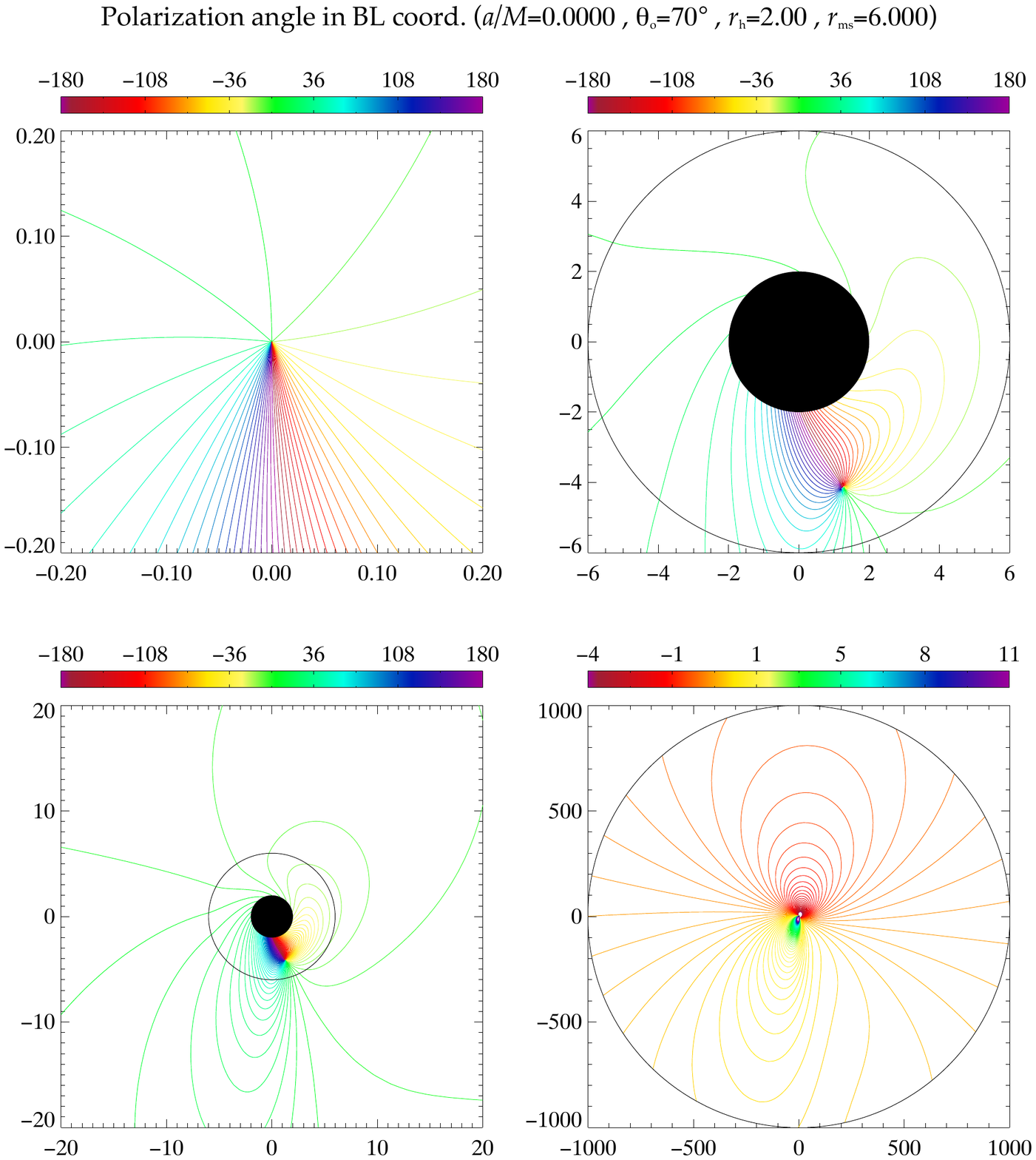}
  \caption{The colour-coded contour plot shows the change of polarization angle $\Psi$ over the equatorial accretion disk plane, as it is detected by a distant observer. The angle $\Psi$ is defined with respect to the direction of polarization vector emitted from an equatorial accretion disk at Keplerian rotation near a non-rotating black hole (spin parameter $a=0$) in Boyer-Lindquist coordinates (BL). Below the ISCO, denoted by a black circle at dimensionless radius $R=6$ (units of gravitational radius $GM_\bullet/c^2$), matter is free-falling inward while conserving its angular momentum content. Four panels are centered on the black hole with a gradually increasing radial resolution (going from the top-left panel to the bottom right one). A critical point is clearly visible, where the emerging polarization angle becomes undefined. $\Psi$ angle wobbles for the photons originating from a source orbiting above the critical point; the polarization angle rotates below the critical point (on top of each panel, see the colour scale in degrees). In the top-left panel the radius of the horizon is rescaled to the origin of the plot, so that the near-horizon region is resolved into maximum detail. Other values of the model parameters are specified above the figure. For further details, see \citet{2005ragt.meet...47D}.}
  \label{fig1}
\end{figure}

Radiation flux $F_{\rm{o}}$ detected by a distant observer at frequency $\nuo$ is
\begin{equation}
 F_{\rm{o}}(\theta_{\rm o},\nu_{\rm o})= \int_{\cal{O}}{\rm d}{\cal{P}}\,I_{\rm{o}}(\theta_{\rm o},\nu_{\rm o}),
 \label{fo}
\end{equation}
where the integration is performed over the detector plane, which is located at a viewing angle $\theta_{\rm o}$ in the asymptotically flat radial infinity, far from the black hole. Integration can be written in terms of the source photosphere, which is connected with the detector by light rays (null geodesics):
\begin{eqnarray}
 F_{\rm{o}}(\theta_{\rm o},\nu_{\rm o})=
 \int_{\cal{S}} & {\rm d}{\cal{S}} & \int_{g_{\rm{min}}}^{g_{\rm{max}}} {\rm d}{g}\,
 I_{\rm{e}}(R_{\rm e},\theta_{\rm e},\nu_{\rm e}) \nonumber \\
 &\times& T(R_{\rm e},\theta_{\rm e},\nu_{\rm e};\theta_{\rm o},\nu_{\rm o}).
 \label{fo1}
\end{eqnarray}
The transfer function $T$ is determined by light-bending effects and it defines the fraction of locally emitted energy that reaches the detector. The locally emitted intensity is related to the observed intensity by $I_{\rm{e}}(R_{\rm e},\nu_{\rm e})/\nu_{\rm e}^3=I_{\rm{o}}(\nu_{\rm o})/\nu_{\rm o}^3$.

Polarimetry reaches further and it can help us to improve the constrains on the model parameters. Different approaches to polarimetry have been developed in the literature \citep{1972NPhS..240..161C,1974ApJ...189...39A,1977MNRAS.179..691P,1980ApJ...235..224C}. In GR, a covariant definition of polarimetric quantities is more natural and it was introduced by various authors \citep{1974CMaPh..38..103M,1975A&A....44..389B,1975ApJ...202..454A}. To this end, \citet{1980RSPSA.370..389B} and \citet{1989rfmw.book.....A} employ the polarization tensor, $J_{\alpha\beta\gamma\delta}\,\equiv\,\frac{1}{2}\langle F_{\alpha\beta}F_{\gamma\delta}\rangle$. Further discussion can be found in \citet{2003MNRAS.342.1280B,2016MNRAS.462..115D,2020MNRAS.491.4807M,2020A&A...641A.126B}. Recently, \citet{2021ApJ...906...34K} developed a versatile formalism to capture various processes of radiation transfer and reprocessing by plasma moving close to Kerr black hole. The method can be adapted to different geometry of plasma streams, both accretion, outflows, and ejection of collimated jets. The general framework is particularly needed in the context of radiative transfer in coronae above the accretion disk because its size and shape are still largely unknown \citep[e.g.][and further references cited therein]{2015MNRAS.451.4375F,2019ApJ...875..148Z,2019Natur.565..198K,2020A&A...644A.132U}. It appears that the X-ray active coronae are formed by hot plasma that is concentrated in a relatively compact region. It is expected that photons that scatter multiple times in the corona are constrained to a plane parallel to the accretion disk, and thus are preferentially polarized in the vertical direction \citep[e.g.][]{2010ApJ...712..908S,2020MNRAS.493.4960T,2021MNRAS.501.3393T}. The more scattering events, the stronger the degree of polarization is expected.

The wave vector of light propagating in curved metric, $k_\alpha\equiv \psi_{,\alpha}$, is parallel transported along null geodesics:
\begin{equation}
k_{\alpha;\beta}\,k^\beta=0,\quad k_\alpha k^\alpha=0,
\end{equation}
whereas the propagation law in the empty space can be written in the form $DF_{\alpha\beta}-2\theta F_{\alpha\beta}=0$ with $\theta\equiv-\frac{1}{2}{k^\alpha}_{;\alpha}$, $D\,\equiv\,u^\alpha\nabla_\alpha$ \citep{1977ApJ...217..353A}. For the emission from a Keplerian disk in the equatorial plane of Kerr metric,
\begin{equation}
1+z=\frac{r^{3/2}-3r^{1/2}+2a}{r^{3/2}+{a}-\xi}, \quad
\cos\vartheta=\frac{g\eta^{1/2}}{r},
\end{equation}
where $\vartheta$ is the local emission angle, $\xi$ and $\eta$ are constants of motion.

Light rays and the associated wave fronts of electromagnetic radiation do not depend on polarization \citep[in the geometrical optics approximation, which we adopt here;][]{Perlick2004}. Equivalence between the formalism of light-rays and wave fronts follows from the Fermat principle. As mentioned above, this brings us to an interesting analogy between the light propagation in a vacuum curved spacetime versus the light bending by material media in a flat spacetime. Light rays in a spherically symmetric and static spacetime can be characterized by an effective index of refraction. For example, in Schwarzschild metric, the effective permeability is $\mu=\sqrt{1-2M/r}$ \citep{1973PhRvD...7.2807M,1977PhRvD..16..933H}. The eikonal equation describing the light propagation in the approach of wave-forms,
\begin{equation}
\!-\left(1-\frac{2M}{r}\right)(\psi_{,r})^2+\left(1-\frac{2M}{r}\right)^{-1}\!\!\!\!(\psi_{,t})^2-r^{-2}(\psi_{,\phi})^2=0,
\end{equation}
is a non-linear partial differential equation. Here it can be solved by separation of variables, $\psi(t,r,\phi)\equiv R(r)+\alpha\phi-\omega t$, where the wave front is found by solving
\begin{equation}
\left(1-\frac{2M}{r}\right)(R^\prime)^2=\left(1-\frac{2M}{r}\right)^{-1}\omega^2-r^{-2}\alpha^2,
\end{equation}
where the prime denotes derivative with respect to $r$. The solution adopts the form $\psi(t_0+n\,{\delta}t,r,\phi)=\psi(t_0,r_0,0).$ 

\begin{figure}[tbh!]
\vspace*{-7em}
  \centering
  \includegraphics[scale=0.45]{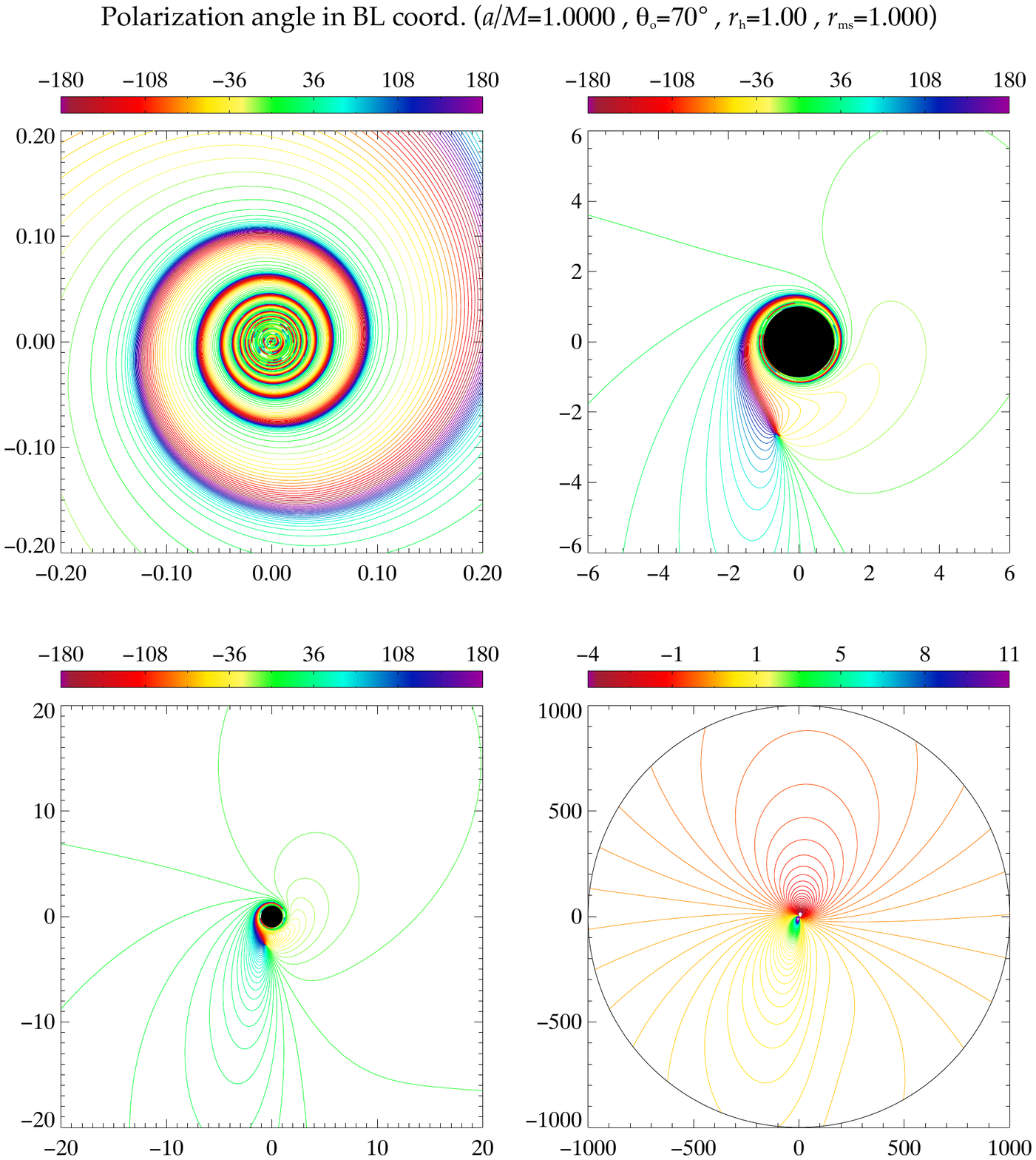}
  \caption{The same as in the previous figure but for a maximally rotating black hole, $a=1$. In this case the ISCO radius coincides with the black hole horizon and the contours are further deformed by the action of frame-dragging. The critical point is seen again near the event horizon, but now it occurs above the ISCO.}
  \label{fig2}
\end{figure}

The wave fronts are perpendicular to light rays and they wrap around the event horizon. Bending of light rays is thus complementary to this wrapping form of the wave fronts. This can be extended to Kerr metric, where the separation of variables and the solution for the eikonal equation follow from Carter's solution of the scalar wave equation, $\psi=R(r)+T(\theta)+\alpha\phi-\omega t$ \citep{1968PhRv..174.1559C,2021arXiv210508556M}. Because of frame-dragging in the ergoregion, the instantaneous shape of wave fronts becomes progressively deformed around the horizon in the sense of spin. In order to disentangle the influence of infinitely growing frame-dragging, we can employ Kerr ingoing coordinates \citep[e.g.][]{2015CQGra..32l4006T,2017grav.book.....M}, which are attached in a natural way to the rotating spacetime.

Let us note that the system of black hole -- orbiting source is highly non-spherical, and an interesting level of polarization can be expected. However, in order to observe these polarization changes to occur on the dynamical time-scale, one has to achieve large S/N ratio, which is a difficult goal for polarization measurements \citep{2015ApJ...807...53I,2020ApJ...889..111A}. Near Kerr black hole the orbital period of matter co-rotating along a circular trajectory of $r=\mbox{const}$ is 
\begin{equation}
T_{\rm{}orb} = 310~\left(r^\frac{3}{2}+a\right)\,M_7\quad\mbox{[sec]}.
\label{torb}
\end{equation}
It appears that the typical resolution of the observation needs to be as good as a few minutes, and the characteristic changes of polarization should be expected over a shorter period. On the other hand, when integrating over a longer duration of the observation, the observed polarization becomes diminished.

For an infinitesimal (point-like) source representing an element on the surface of an accretion disk, a distant observer measures the flux entering the solid angle ${\rm d}\Omega_{\rm o}$ that can be associated with an area element on the observer's detector, ${\rm d}S_{\rm o}{\equiv}D^2\,{\rm d}\Omega_{\rm o}$, where $D$ is the luminosity distance between the observer and the source. The photon flux received by the detector is then given by \citep{2005ragt.meet...47D}
\begin{eqnarray}
\label{flux}
N^{S}_{\rm{}o}(E)\equiv\frac{{\rm d}n(E)}{{\rm d}t\,{\rm d}S_{\rm o}}
&=&{\int}{\rm d}\Omega\,N_{\rm loc}(E/g)\,g^2\,,
\\
N_{\rm loc}(E_{\rm loc})&\equiv&
\frac{{\rm d}n_{\rm loc}(E_{\rm loc})}{{\rm d}\tau\,{\rm d}S_{\rm loc}\;
{\rm d}\Omega_{\rm loc}}
\end{eqnarray}
$N_{\rm loc}$ is the locally emitted flux emerging from the surface of the disk, ${\rm d}n(E)$ is the number of photons with energy in the interval ${\langle}E,E+{\rm d}E\,\rangle$ and $g=E/E_{\rm loc}$ is the combined (gravitational and Doppler) shift of photon energy.

\section{Polarization angle near black hole}
The polarization angle is sensitive to the effect of rotation upon propagation in a (vacuum) curved spacetime, however, the degree of polarization is not influenced by strong gravity. To constrain the polarization angle changes by measurement, the major practical difficulty arises from the need to acquire sufficient signal-to-noise ratio, and so a large collecting area of the upcoming instruments is required \citep{2016SPIE.9905E..1QZ,2021SPIE11444E..2XM}. This is necessary in order to be able to determine geometry of the system and to break degeneracies among free parameters that are inherent to accretion models. Different physical effects influence the polarization of light as it propagates towards an observer. In our idealized graphs, we examine only the effect of the gravitational field. The change of the polarization angle is defined as the angle by which a vector parallelly transported along the light geodesic rotates with respect to a selected physical frame (let us note that in vacuum the polarization pseudo-vector is parallelly transported along the null geodesic of the photon ray). 

\begin{figure}[tbh!]
\vspace*{-7em}
  \centering
  \includegraphics[scale=0.45]{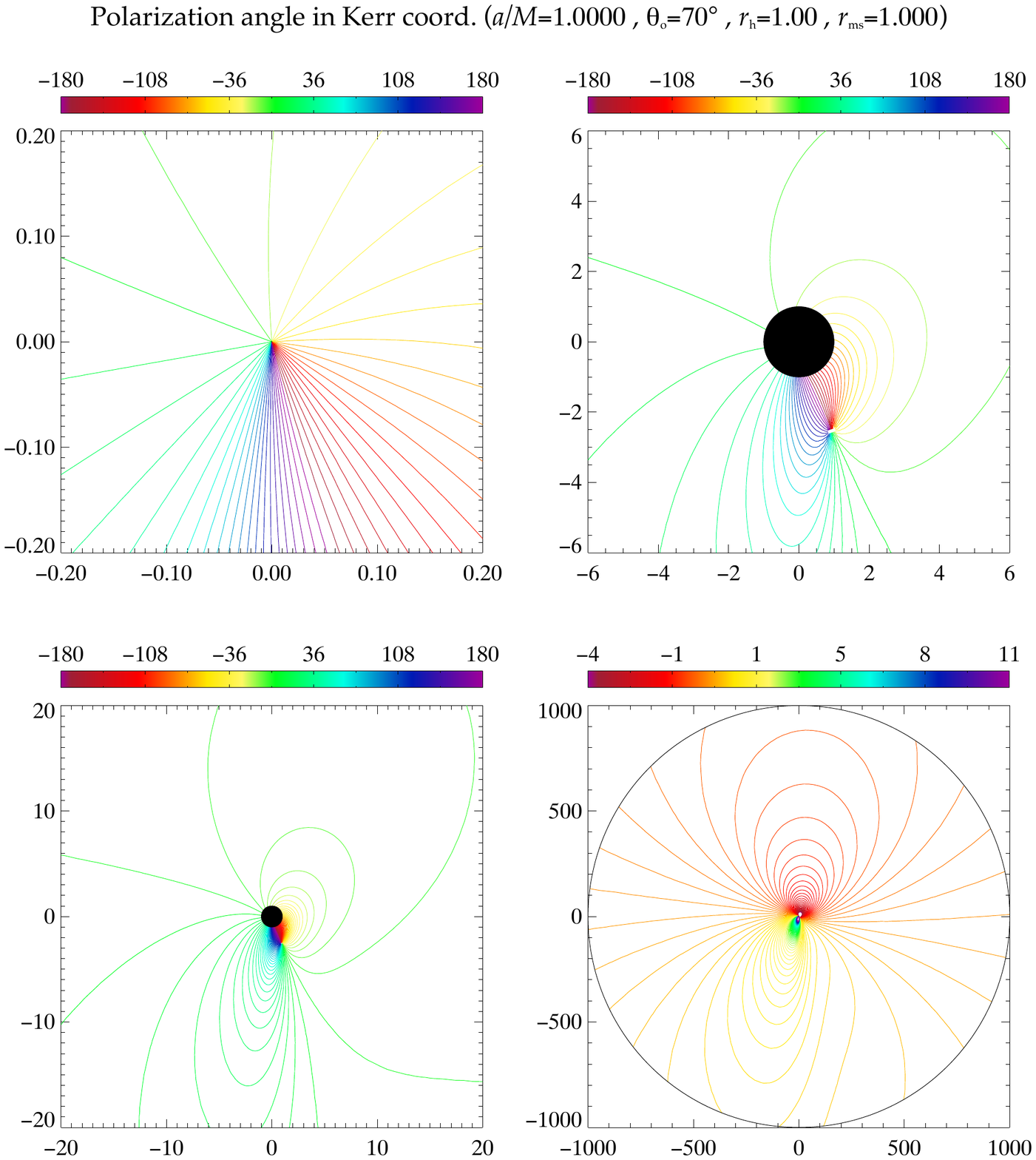}
  \caption{As in the previous figure but with contours constructed with respect to Kerr ingoing coordinates. In this case the effect of frame-dragging is absorbed into the definition of coordinates that are attached to ingoing photons and, therefore, they rotate progressively faster at small radii.}
  \label{fig3}
\end{figure}

The graphical representation of the observable (theoretical) changes of the polarization angle is shown in Figure~\ref{fig1}. Four frames of contour plots correspond to different range of radius and a gradually increasing resolution on the central region of Schwarzschild black metric, where the gravitational effects are most prominent. This figure captures the equatorial plane for given values of spin $a$ and observer's view angle $\theta_{\rm o}$ (almost edge-on view of the equatorial plane is shown, i.e.\ almost perpendicular to the rotation axis). The radius extends up to $r=10^3$ gravitational radii ($R_{\rm g}$). In a non-rotating case, the black hole is static (spin parameter $a=0$), and so there is no effect of relativistic frame-dragging, although matter is in Keplerian orbital motion. This can be compared with Figure~\ref{fig2}, where we show the opposite extreme of a maximally rotating black hole ($a=1$). In the latter case the critical point occurs in the region of stable orbits above the ISCO. In order to eliminate the effect of frame dragging in the plot, the local frame can be attached to infalling photons, which also feel the dragging effect. The resulting shape of the contour lines becomes less cumbersome. Figure~\ref{fig3} demonstrates the situation for $a=1$; although this is actually the same state of extreme rotation as in the previous figure, the use of ingoing coordinates helps us to clear the graph from the effect of dragging. 

In NIR/sub-mm and X-ray domains we expect the synchrotron or synchrotron-self-Compton mechanisms as the dominant process generating the signal from relativistic electrons gyrating around magnetic-field lines in the accretion flow \citep{1986rpa..book.....R}. The energy distribution of synchrotron electrons can be approximated by a power-law 
\begin{equation}
N(\gamma)= N_0\gamma^{-s}, \quad \gamma \leq \gamma_c,\quad s={\rm const},
\label{elec_dist}
\end{equation}
where $N(\gamma), \gamma$ and  $\gamma_c$ are the electron distribution function, Lorentz factor of the electrons, and the cut-off energy, respectively \citep{2011MNRAS.413..322Z}. 

The volume spectral emissivity $\eta(\nu)$ of relativistic electrons with a power-law energy distribution is given by integration,
\begin{equation}
 \eta(\nu)  \propto \int n(\gamma)\; {{\cal P}}(\nu)\;d\gamma
       \;\propto\; B^2\int\gamma^{2-s}\;\delta(\nu-\nu_{\rm c})\;d\gamma,
       \label{semis}
\end{equation}
which gives $\eta=B^{1+\alpha_{s}}\nu^{-\alpha_{s}}$ with the spectral index $\alpha_{s}\equiv(s-1)/2$. The linear polarization is  $Q=[(s+1)/(s+7/3)]\,I$. The synchrotron radiation is also circularly polarized, $V=3\gamma^{-1}I$, and it can be self-absorbed in the source. 

The cyclotron radiation of a single gyrating particle is circularly polarized in the direction of magnetic vector and linearly polarized in the perpendicular direction. The polarization produced by particles with an isotropic distribution is almost completely linear; only highly anisotropic distributions exhibit some degree of ellipticity in polarization. The circular polarization typically decreases towards short mm-wavelengths \citep{2003ApJ...588..331B} but the mechanism of conversion has been examined in GR regime \citep{2021arXiv210300267M,2021arXiv210509440G,2021MNRAS.505..523R}. Furthermore, in a situation when synchrotron radiation is significantly inverse-Compton scattered, the resulting spectrum is described as a synchrotron--self-Compton (SSC) process. 

In plasma, the circular polarization can be produced not only via intrinsic emission but also by the Faraday conversion in the course of propagation. By measuring the polarization we can assess the importance of different emission processes. Other aspects of General Relativity with polarimetry were examined by \citet[][case of accreting pulsars]{2004A&A...426..985V}, \citet[][Compton reflection from an X-ray-illuminated accretion disk]{2004MNRAS.355.1005D}, and \citet[][polarization by scattering on infalling clouds]{2006MNRAS.365..813H,2006PASJ...58..203H}. In the near-horizon region (typically up to $\simeq10$--$15\,R_{\rm g}$), the Faraday rotation of a plasma filled magnetosphere and birefringence of the Quantum Electrodynamics vacuum can impact the observable polarization \citep{2021ApJ...914...51K}. The relative importance of both effects depends on the magnetic field and the photon frequency. The role of vacuum birefringence depends on the strength of the magnetic field in units of the critical magnetic field, $B_{\rm c}=m_{\rm e}^2c^3/e\hbar\simeq 4.4\times10^{13}$G; hence it is relevant in intense magnetic fields especially near magnetized compact stars.

The above-mentioned behaviour of the polarization angle at certain radius close to the black hole horizon is a signature of strong gravity. Although this provides a specific evidence which in future can help us to measure the model parameters, so far its practical use has been attempted only in a few preliminary cases with tentative results, such as the SMBH in Sagittarius A*, where the near-infrared polarimetry has been achieved from ground-based observations \citep{2006A&A...458L..25M, 2015A&A...576A..20S,2020A&A...643A..56G}, and M87 nucleus with the EHT in mm domain \citep{2020A&A...637L...6K}. Even though these constraints have had significant uncertainties, it is important to note that the synergy between space based (X-ray) and ground based (NIR, mm) techniques will be essential to explore the inner regions near the event horizon. 

\begin{figure}[tbh!]
\vspace*{-9em}
  \centering
  \includegraphics[width=0.49\textwidth]{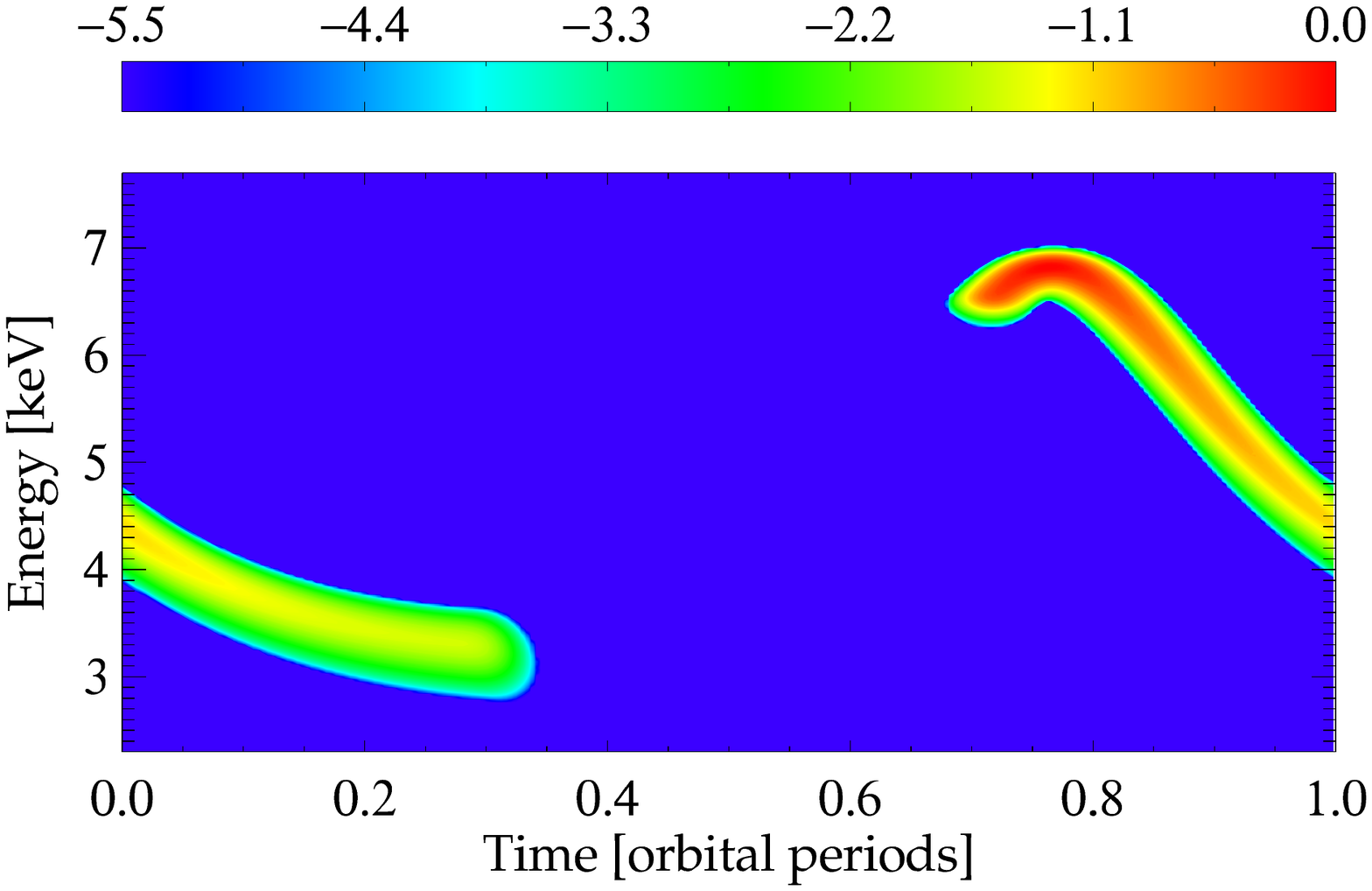}
  \vspace*{-12em}
  \caption{The effect of obscuration on the expected dynamical spectrum from a small spot located on the surface of an equatorial accretion disk at radius $r= 5GM/c^2$. As a model for the relativistic spectral line broadening, the intrinsic energy of the spectral line is set at $6.4$ keV. Energy and radiation flux changes are caused by the combined influence of the orbital motion, the gravitational redshift and the light focusing. The predicted photon flux is colour-coded in a logarithmic scale. The flux drops when the self-eclipse occurs between the azimuthal angles $315\leq\phi\leq540$ deg, as measured from the point of the closest approach to the observer (view angle inclination $45$ deg). The case of a maximally rotating Kerr black hole is assumed \citep[arbitrary units;][]{2004ApJS..153..205D}.}
  \label{fig5}
\end{figure}

The outlined concept becomes less conclusive and its interpretation more cumbersome once the effects of magnetic fields and the geometrical thickness or a tilt of the accretion flow start to play a significant role. Firstly, the elevation of the accretion disk surface above the mid-plane can lead to self-obscuration of some regions \citep[those that are shielded from the direct view, especially in the case of a large inclination angle; e.g.][]{1992ApJ...400..163B,1992A&A...257..531K,1998MNRAS.301..721U,2007A&A...474...55F}. Part of the signal may thus be missing compared to the complete un-obscured geometry without the influence of self-eclipses (fig.\ \ref{fig5}). Secondly, the dynamical role of the magnetic field can change the inner boundary condition for the torque acting on the inflowing material, and hence this can shift the radial position of the inner edge of the accretion flow away from the ISCO. The accretion disk can extend further down into the plunging region, or vice versa it can be truncated before reaching the ISCO by the enhanced magnetic pressure \citep[][and further references cited therein]{2008ApJ...675.1048R,2018ApJ...855..120T,2003PASJ...55L..69N,2016MNRAS.462..636A}. 

A toy model for polarized images of synchrotron emission from an equatorial source near a Kerr black hole by using a semi-analytic solution has been recently developed by \citet{2021arXiv210509440G}. Despite simplifying assumptions that are needed to tackle the problem, their approach can be directly compared to simulated and observed values of near-infrared flares, where the effects of Faraday rotation, absorption, and background emission can be neglected. In agreement with previous results of various authors, the latter paper shows that while the geometrical effect of black-hole spin on photon trajectories is only moderately visible in the outgoing electromagnetic signal, the spin can strongly influence the accretion dynamics and the emissivity profile of plasma close to the event horizon. This corresponds to the fact that the location of the innermost stable circular orbit is rather sensitive to angular momentum of the black hole rotation, which then determines the truncation radius of the accretion disk, where most of the emerging signal originates.

Let us note that the additional parameters can lead to a certain degeneracy of the model parameters and subsequent difficulties in converging the procedure to fit their values. This may be particularly serious limitation for polarimetry technique, which is known to be sensitive to the assumptions about the geometry of the system. For example, a self-eclipse of the portion of the accretion flow will cause the effect not to be revealed due to the lack of data. Apart from the geometrical rotation of the polarization angle, which has been of our main interest in this paper, the relativistic aberration also significantly influences the emerging polarization. Despite the challenge it turns out that polarimetry in submillimeter band has been successful in putting first constrains on the accretion flow properties in the Galactic center and in M87 galaxy, with some prospects for additional suitable sources in the future \citep[][and further references cited therein]{2021ApJ...910L..14G,2021arXiv210300267M}.

\section{Strong-gravity effects on the appearance of stars and hot spots near SMBH}

In \citet{1972ApJ...173L.137C}, the authors studied the effect of strong gravity on a star orbiting at a small distance (namely, $20$ and $3$ gravitational radii in their examples) from a maximally rotating black hole. The original motivation of the motion of a bright object has mostly been employed in terms of gas orbiting in the form of a gaseous disk with hot spots, spiral waves, vortices, and shocks \citep{1992Natur.356...41A}, which may be relevant for a large fraction of active galactic nuclei in general to interpret their variability amplitudes and timescales. However, it turns out that our own Galactic center (Sgr~A*), which is extremely underluminous with the Eddington ratio of $\eta \lesssim 2 \times 10^{-8}$ \citep{1998ApJ...492..554N}, is a very suitable target where the motion of individual stars and gaseous clumps can be resolved during their approach to the pericenter and the passage at a small distance from the central black hole. The pericenter passage of the S2 star at $\sim 119\,{\rm AU}$ from Sgr~A* was closely monitored in May 2018 \citep{2018A&A...615L..15G,2019Sci...365..664D}, and the pericenter transit of the G2 source in $2014.2$ at $\sim 135$--$185$ AU was intensively observed to look for potential tidal effects and the disruption of the source \citep{2013ApJ...774...44G,2013ApJ...773L..13P,2015ApJ...800..125V}. The peribothron of G1 in $2001.3$ at $\sim 292\,{\rm AU}$ was analyzed in more detail recently, with the apparent drop in the $L$-band flux density  interpreted as a signature of tidal truncation during the pericenter phase of the orbit \citep{2017ApJ...847...80W}. 

The monitoring of individual stars close to the Sgr~A* has revealed sources with the pericenter distance of the order of $1000$ gravitational radii \citep[$\sim 2953$ gravitational radii for the S2 star;][]{2018A&A...615L..15G,2019Sci...365..664D} and recently even one order of magnitude closer \citep[$\sim 440$ gravitational radii for S62 and $\sim 312$ gravitational radii for S4714;][]{2020ApJ...889...61P,2020ApJ...899...50P}. Hence, the original motivation of Cunningham and Bardeen will likely be of direct relevance, especially once $30+$ meter class telescopes come in operation that will be able to detect faint stars at a distance of only $100$ gravitational radii and less from Sgr~A*. Statistically speaking, using a general power-law distribution of a stellar density in the Nuclear Star Cluster (NSC), one obtains the number of stars on these length-scales less than unity \citep{2018AN....339..324Z}. On the other hand, scattering events in the NSC are expected to bring stars occasionally on highly eccentric orbits with pericenter distances in this sparse region \citep{2017ARA&A..55...17A,2020arXiv201103059A}.

\begin{figure}[tbh!]
    \centering
    \includegraphics[width=\columnwidth]{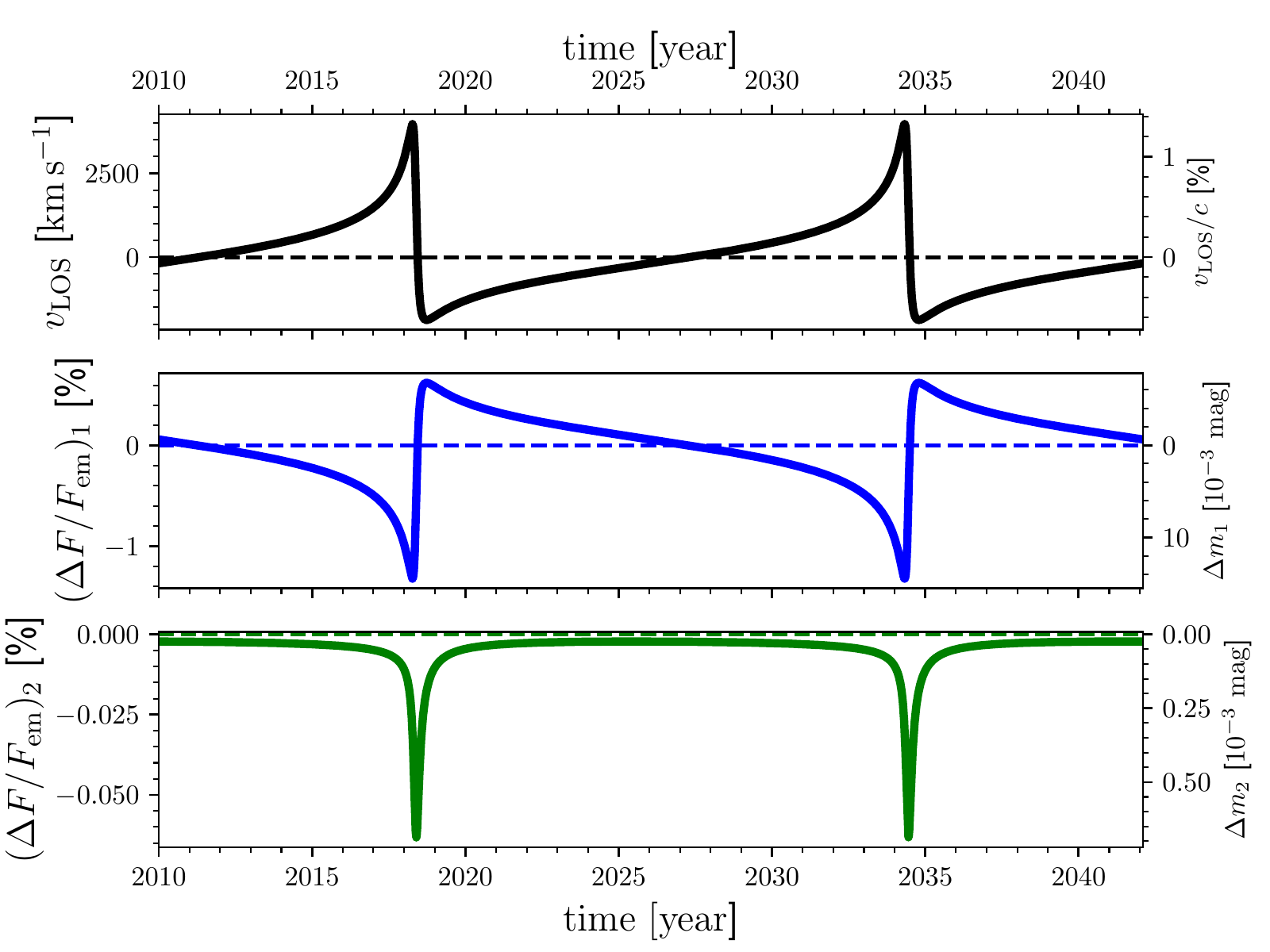}
    \caption{Doppler boosting for the S2 star along its elliptical orbit around Sgr~A*. Top panel: The line-of-sight velocity in km/s as a function of time. The $y$-axis on the right stands for the fraction of the light speed in percent. Middle panel: The relative Doppler boosting change in percent including $\mathcal{O}(v/c)$ terms (nonrelativistic). The $y$-axis on the right expresses the change in $10^{-3}$ mag. Bottom panel: The relative Doppler boosting change in percent due to $\mathcal{O}(v^2/c^2)$ terms (gravitational redshift and transverse Doppler shift). The $y$-axis on the right expresses the change in $10^{-3}$ mag.}
    \label{fig_doppler_boosting_S2}
\end{figure}

For the S2 star, two general relativistic effects were already detected with a high significance, namely the combined gravitational and transverse Doppler shift thanks to absorption lines,
$z=\Delta \lambda/\lambda\approx 200$ km\,s$^{-1}/c$ \citep[][]{2018A&A...615L..15G,2019Sci...365..664D} and the prograde pericenter (Schwarzschild) precession of its orbit \citep[$\delta \phi \approx 12'$ per orbital period;][]{2020A&A...636L...5G}; see also \citet{2017ApJ...845...22P} for the first indication of the prograde general relativistic pericenter shift. S-stars have also been of interest in terms of their lensing images, wih Sgr~A* acting as a giant lens for bending stellar photons. The secondary image is the brightest one, with higher-order lensing images becoming progressively fainter and positioned closer to the apparent horizon. When one considers the best-fit orbits of S stars, it is possible to predict and calculate the peak magnitude and the position of the secondary image. For the best configurations, the secondary image has the lowest $K$-band magnitude of $\sim 20$ and the separation from Sgr~A* of a few $0.1$ mas at most \citep{2005ApJ...627..790B,2009ApJ...696..701B}. This is still undetectable even for the current Very Large Telescope second-generation interferometer GRAVITY \citep{2008SPIE.7013E..2AE}. Such faint images are still below its sensitivity limit of $K\sim 19$ in the astrometry mode with the spatial resolution of $\sim 10\mu{\rm as}$.

Another effect that is expected to leave distinct imprints on the continuum flux density of stars orbiting close to the supermassive black hole is the Doppler boosting and beaming \citep{1996ApJ...470..743K,2005A&A...441..855P,2006ApJ...639L..21Z,2007ApJ...670.1326Z,2020ApJ...905L..35R}. The S-cluster stars orbit the supermassive black hole on the scales from several thousand to hundred gravitational radii and reach velocities up to $10\%$ of the light speed \citep{2020ApJ...899...50P}; their flux as detected by a distant observer on the Earth is thus modified by the above-mentioned combination of special and general relativistic effects, namely time dilation, light aberration, and the photon frequency shift due to the motion and gravitational redshift. All these effects lead to the Doppler boosting of the observed flux in the near-infrared bands. The Doppler boosting of the continuum flux is only moderate, at the level of the observational uncertainties of current instruments, but for S stars with tight orbits it can lead to the relative change of a few percent. Specifically, \citet{2020ApJ...905L..35R} estimates the relative change of the photometric signal due to the Doppler boosting at $\sim 6\%$ for the fastest candidate stars, S62 and S4714, and $\sim 2\%$ for S2. 

In Fig.~\ref{fig_doppler_boosting_S2}, we show the line-of-sight velocity in the top panel, the relative change of the photometric signal due to $\mathcal{O}(v/c)$ terms (nonrelativistic Doppler boosting; middle panel), and the relative change of the continuum flux due to $\mathcal{O}(v^2/c^2)$ terms (general relativistic gravitational redshift and special relativistic transverse Doppler shift; bottom panel). The total relative change of the near-infrared flux density of S2 due to $\mathcal{O}(v/c)$ (nonrelativistic) Doppler boosting, denoted as $(\Delta F/F_{\rm em})_1$, is $\sim 1.95\%$ or $\sim 21\times 10^{-3}$ magnitudes. This order of magnitude difference in flux density will be detectable with the future near- and mid-infrared instruments at Extremely Large Telescope, whose photometric sensitivity is expected to be at $10^{-3}$ mag level \citep{2020A&A...644A.105H}. On the other hand, the total relative flux change due to $\mathcal{O}(v^2/c^2)$ terms, denoted as $(\Delta F/F_{\rm em})_2$, which corresponds to special and general relativistic Doppler boosting, is at the level of $\sim 0.06\%$ or $0.7\times 10^{-3}$ magnitudes. However, taking into account the crowded stellar field in the Galactic center, in particular the overlapping point spread functions of fast-moving S stars \citep{2012A&A...545A..70S,2020ApJ...905L..35R}, this amount of the relative change will be very challenging even with upcoming instruments.     

\begin{figure}[tbh!]
    \centering
    \includegraphics[width=\columnwidth]{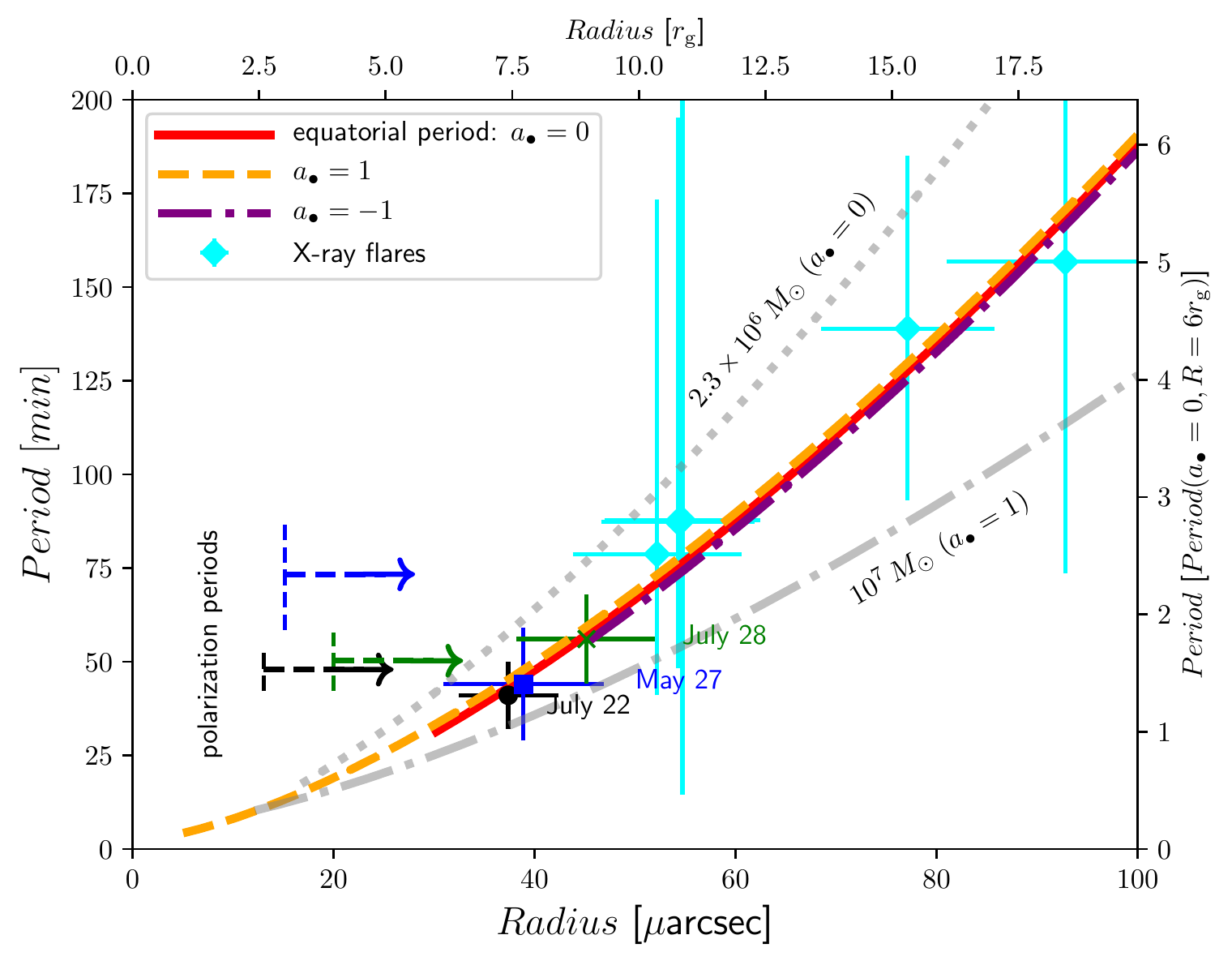}
    \caption{Period--radius relation for the three near-infrared flares whose clockwise orbital motion was detected by GRAVITY in May--July 2018 \citep{2018A&A...618L..10G} to which five X-ray events are added from \citet{2017MNRAS.472.4422K}. The position of the flares in this plot is consistent with the orbital motion in the equatorial plane near the ISCO of $\sim 4\times 10^6\,M_{\odot}$ black hole. Even when uncertainties are taken into account, the detection of the bound motion on the scales of several tens of microarcseconds implies the presence of the compact mass of $10^6-10^7\,M_{\odot}$; see the gray dotted and dash-dotted lines. The period of $45\pm 15$ min for the near-infrared flares is comparable to the period of the detected rotation of the polarization angle. The upper $x$-axis is expressed in gravitational radii of $4.14\times 10^6\,M_{\odot}$ black hole. The right $y$-axis is expressed in the equatorial orbital period for a non-rotating black hole of $4.14\times 10^6\,M_{\odot}$ at the ISCO. The data for the three near-infrared flares, including the polarization periods, were adopted from \citet{2018A&A...618L..10G}, to which we have added five X-ray flares from \citet{2017MNRAS.472.4422K}; the latter ones (shown in the light-blue color) occur at a larger distance and they have significantly larger error bars, nonetheless, they still remain consistent with the expected period--radius curve.}
    \label{fig_period_radius_hotspot}
\end{figure}

Apart from fast-moving stars on elliptical orbits, Sgr~A* is the only supermassive black hole, around which one can monitor orbiting bright spots close to the ISCO \citep{2018A&A...618L..10G} using the GRAVITY Very Large Telescope Interferometer. 
Hot spots or orbiting bright localized emission regions in the accretion flow were frequently employed as a useful toy model to explain the variability in galactic nuclei, specifically bright flares \citep{1972ApJ...173L.137C,1991A&A...245..454A,1992ApJ...400..163B,1992A&A...257..594B,1994ApJ...425...63B,2004A&A...420....1C,2008A&A...487..815P,2013A&A...556A..77P}, including both the total and polarized flare emission. The model of the hot spot orbital motion close to the ISCO of $4\times 10^6\,M_{\odot}$ supermassive black hole was elegant in explaining $\sim 1$ hour lasting bright states of Sgr~A*, whose duration is comparable to the orbital period of hot spots; see Fig.~\ref{fig_period_radius_hotspot} for the general period--radius relation. In principle, by comparing near-infrared and X-ray light curves with the synthetic light curves produced by an orbiting hot spot, it is possible to infer Sgr~A* mass, spin, and the viewing angle \citep{2005MNRAS.363..353B,2006ApJ...636L.109B,2006A&A...458L..25M,2010A&A...510A...3Z}. 

The special and general relativistic effects that modulate the emission detected by a distant observer are Doppler boosting, lensing, gravitational redshift, light-travel time delay, and the change of the polarization angle. Selection biases (sensitivity of the detectors and the minimum duration of the observational runs) likely influence the frequency of detections of flares at different energy, where the X-ray events occur somewhat farther out from the central black hole. In this context it is interesting to note that magnetic reconnection rate could become enhanced closer to the event horizon, where also the tidal effects are stronger. In combination of the intrinsic physical and observational effects this may imply that the near-horizon blobs are produced with shorter duty rates and smaller sizes compared to distances at tens--hudreds of gravitational radii \citep{2020MNRAS.499.1561Z,2020arXiv200514251B}. This further points to the importance of the black hole ergosphere near a rapidly rotating black hole in enhancing the magnetic reconnections in the accreting plasma, which is relevant when the magnetic field is anchored in the central black hole and the accretion flow, and is used to extract its rotational energy \citep{2008ApJ...682.1124K,2019MNRAS.487.4114Y,2012CQGra..29c5010K,2020arXiv201215105K,2021PhRvD.103b3014C}.

Some bright X-ray flares exhibit a peak-shoulder structure, where the narrower peak is due to lensing, and hence a larger solid angle, and a broader sub-peak due to Doppler boosting \citep{2017MNRAS.472.4422K}. This structure is particularly visible for hot spots orbiting on nearly edge-on orbits. The five bright X-ray flares analyzed by \citet{2017MNRAS.472.4422K}, which were previously reported by \citet{2001Natur.413...45B}, \citet{2003A&A...407L..17P}, \citet{2008A&A...488..549P}, \citet{2012ApJ...759...95N}, and \citet{2015A&A...573A..46M}, are consistent with the orbital motion at the distance of $\sim 10$--$20$ gravitational radii, and a moderate view angle.

\citet{2018A&A...618L..10G} detected  closed, clockwise orbital motions of emission centroids during three flares in May-July 2018. With the inferred orbital period of $45\pm 15$ minutes and the distance of $\sim 6$--$10$ gravitational radii, the flares are consistent with the hot spot orbiting close to the ISCO of the Schwarzschild or Kerr black hole of $4.14\times 10^6\,M_{\odot}$ (see Fig.~\ref{fig_period_radius_hotspot}), and hence closer than the X-ray flares. The flux modulation during the flares is moderate and it implies near to face-on orbits. For both X-ray and near-infrared flares, they stay luminous only for a single orbital timescale at most at the corresponding inferred distance. For X-ray flares, the enhanced emission lasts close to half of the orbital timescale. Hence, these are transient phenomena, potentially related to the clumpy accretion and/or magnetic reconnection in a hot accretion flow \citep{2014ARA&A..52..529Y}. 

A number of bow shock sources were detected in the innermost parsec of the Galactic centre \citep{2010A&A...521A..13M,2021ApJ...909...62P}. They exhibit distinct characteristics, including an excess towards mid-infrared wavelengths and a significant linear polarization. These features indicate the presence of a non-spherical dusty envelope. The Dusty S-cluster Object shows similar characteristics and it is a candidate for the closest bow shock with a detected proper motion in the vicinity of Sgr A*, with the pericentre distance of only $\sim2000$ Schwarzschild radii. Because of the alternative and quite likely interpretation in terms of a dust-enshrouded stellar object, the source has been referred as G2/DSO in our papers  \citep{2015ApJ...800..125V,2017bhns.work..237Z,2020A&A...634A..35P}.

Furthermore, in the Galactic center region, there are indications of relatively intense magnetization of the environment at both larger and smaller scales around the Sgr~A* supermassive black hole. In particular, the multi-wavelength measurements of the Faraday rotation near the Galactic center magnetar \citep{2013Natur.501..391E} revealed the magnetic field of $B\gtrsim 50\mu$G at the projected distance scale of only $3^{\prime\prime}$, corresponding to $0.12$ pc, implying the importance of polarization data. On smaller scales, close to the ISCO, the flare emission is consistent with the magnetic field intensity of $\sim 10$--$100$ Gauss \citep{2012A&A...537A..52E,2020arXiv201109582W}. In addition, the GRAVITY observations detected the polarization angle rotation with the comparable period as the timescale of the orbital motion \citep{2018A&A...618L..10G}. The polarization loop is caused by light bending. This is consistent with the synchrotron-emitting hot spots moving in the ordered, large-scale poloidal magnetic field. The ordered, poloidal magnetic field is also expected to lead to the charging of Sgr~A* to a small value \citep{1974PhRvD..10.1680W,2018MNRAS.480.4408Z} and the Lorentz force acting on flare components \citep{2020ApJ...897...99T}. In this context it is relevant to note that magnetic null points can be identified as possible sites of magnetic reconnection and particle acceleration due to the gravito-magnetically induced electric field within the magneto-ergosphere, i.e., very close to the event horizon \citep{2013IJAA....3...18K}.

As mentioned above, the polarization properties provide important constraints on the geometry of accretion flows in the inner regions of radio galaxies, quasars, and Active Galactic Nuclei (AGN) in general \citep[e.g.][]{1990MNRAS.242..560L,1996ApJ...472..502A,1999AdSpR..23..871K,2021MNRAS.501.3393T}. Four Stokes parameters can extend the traditional spectroscopy especially where the source is non-spherical (namely, accretion disks and jets). In different spectral bands, the signal is sensitive to the emission mechanism in place. A unique possibility has emerged to observe the obscured Broad Line Region in the scattered (off-axis, polarized) light \citep{1993ARA&A..31..473A,2014AdSpR..54.1341G}. However, in the case of extragalactic sources, a limited resolution and S/N ratio severely reduce the confidence of the polarimetrical modelling because several emission regions contribute to the observed signal, namely, the inner accretion disk, the outer dusty torus, an outgoing jet, and the interstellar medium along the line of sight. 

In case of the polarized BLR emission, equatorial as well polar scatterers are of relevance \citep{2020A&A...637A..88R}. Hence different mechanisms form the polarization from the independent areas. The impact of dust grains filling the region is sensitive to the large scale magnetization; whereas in general the resulting polarization degree tends to be suppressed, the magnetically aligned grains can contribute to polarization enhancement. On the other hand, in several cases the near-horizon region has already been resolved with NIR polarimetry \citep{2016A&A...593A.131S,2021MNRAS.500.4319S} and VLBI radio techniques \citep{2020A&A...637L...6K,2021A&A...646A..52M,2021ApJ...910L..14G}.

\section{Conclusions}
Electromagnetic radiation from the inner regions of black hole accretion disks is influenced by strong-gravity effects. As a result of this, the emerging signal bears imprints of the gravitational field structure of the central object to a distant observer. However, the set of possible geometries and physical conditions is extremely broad and it reaches across different wavelengths; in this brief contribution we could not touch all aspects. Instead, we concentrated our attention on X-rays arising very close to the event horizon of accreting SMBHs, where high-energy photons dominate. For  several decades the energetic photons have provided the classical messengers that convey most of our knowledge about black holes. Spectral and timing synergy is particularly successful in constraining the system parameters. 

Even in vacuum the angle of polarization rotates upon light propagation near a black hole. A critical point emerges specifically due to the effect of light bending, and it results in a transition from wobbling to continuous rotation of $\Psi$ at a certain distance of the source. The polarization properties have been employed in NIR spectral band from ground based observations, but not yet in the X-ray domain to constrain the parameters of the central black hole; this should be possible with several upcoming satellite missions in the near future \citep{2019arXiv190409313K,2019SCPMA..6229502Z,2020SPIE11444E..2EC}. Much as interesting the theoretical possibility appears to be, its practical use will be very challenging because the effect of polarization changes tends to reduce the overall polarization degree of the source in comparison with the case when strong gravity is not present. Ambitious proposals to deploy an X-ray instrument with a very large collecting area would be very helpful, but this is an outlook to distant future. See {\sl https://astro.cas.cz/karas/papers/ky/tables.htm} for examples with different configurations. Here one can also explore other strong-gravity effects on the electromagnetic signal over the parameter space.

In summary, given the predominantly planar geometry of an accretion disk and the intrinsic mechanism that polarizes the emerging photons, the directional change (wobbling or rotation) can inform us about the radius near the event horizon where the most of the observed signal arises. On the other hand, depolarization may occur if the source structure is complex or if the signal passes through complex plasma environment. In order to disentangle different counter-acting effects it will be fruitful to perform measurements sensitive over a broad range of wavelength. Further progress will require to differentiate better the signal from the inner accretion flow against the background contamination. A leading message that this report has attempted to convey is the crucial role of synergy between space-borne and ground-based techniques, and the awareness of the importance of the multi-wavelength approach. This should be apparent from the fact that the signal from the inner accretion disk and the plunging region create detectable signatures in all energy bands of the emerging electromagnetic spectrum.

\section*{Acknowledgments}
Continued support is acknowledged from the Czech Science Foundation grant ``Accreting Black Holes in the New Era of X-ray Polarimetry Missions'' (MD, No.\ 21-06825X). We also thank the Czech-Polish mobility program (MZ, M\v{S}MT No.\ 8J20PL037; NAWA No.\ PPN/BCZ/2019/1/00069; NAWA No.\ PPN/WYM/2019/1/00064), the Polish National Science Center (MZ, No.\,2017/26/A/ST9/00756, Maestro 9), and the Czech PRODEX program by ESA (VK, No.\ 4000132152).

\bibliographystyle{model5-names}
\biboptions{authoryear}
\bibliography{vkaras}

\end{document}